\documentclass[draftcls,onecolumn]{IEEEtran}

\usepackage{epsfig}

\begin{document}
\title{Ultra-low-voltage bilayer graphene tunnel FET}

\author{\normalsize Gianluca Fiori, Giuseppe Iannaccone\\
 Dipartimento di Ingegneria dell'Informazione~: Elettronica, Informatica, Telecomunicazioni,\\
Universit\`a di Pisa, Via Caruso, 56126 Pisa, Italy.\\
email : gfiori@mercurio.iet.unipi.it; Tel. +39 050 2217596; Fax : + 39 050 2217522\\}
\maketitle

\bibliographystyle{ieee}

\newpage
\begin{abstract}

In this work, we propose the Bilayer Graphene Tunnel Field Effect Transistor (BG-TFET) as a device suitable for 
fabrication and circuit integration with present-day technology. It provides  high $I_{\rm on}/I_{\rm off}$ ratio at ultra-low supply voltage, 
without the limitations in terms of prohibitive lithography and patterning requirements for circuit integration of 
graphene nanoribbons. 
Our investigation is based on the solution of the coupled Poisson and Schr\"odinger equations in 
three dimensions, within the Non-Equilibrium Green’s (NEGF) formalism on a Tight Binding Hamiltonian. We show 
that the small achievable gap of only few hundreds meV is still enough for promising TFET operation, providing a 
large $I_{\rm on}/I_{\rm off}$ ratio in excess of 10$^3$ even for a supply voltage of only 0.1 V. Key to this performance is the low quantum 
capacitance of bilayer graphene, which permits to obtain an extremely small sub-threshold 
swing S smaller than 20 mV/decade at room temperature.

\end{abstract}

\newpage

\section{Introduction}

Very recent theoretical and experimental papers~\cite{Mccan1,Nilsson,Ohta,Ryzhii} have
shown the possibility of inducing an energy gap ($E_{\rm gap}$) in bilayer graphene by means of an applied 
electric field perpendicular to the graphene plane. This property could open the possibility of fabricating
carbon-based electron devices, for which a semiconducting gap is required,
with state-of-the-art lithography. 
Let us stress that the alternative option of inducing a gap by means of lateral 
confinement to fabricate graphene nanoribbon Field Effect Transistors (FETs)~\cite{Avouris} requires single-atom 
control and prohibitive lithography to define widths close to 1~nm~\cite{fiori_edl_2007}.

Since the largest attainable $E_{\rm gap}$ in bilayer graphene is of 
few hundreds meV, band-to-band tunneling strongly limits device performance, so that the $I_{\rm on}/I_{\rm off}$ 
ratio of the drain current is typically smaller than 10~\cite{Fiori}.

One can however turn this limitation into an advantage, fully exploiting band-to-band tunneling 
instead of avoiding it, like in Tunnel FETs (TFETs), which
 have been widely investigated in the past through experiments and simulations,
considering different channel materials~\cite{Hansch,Koswatta,Jing}.

The main appeal of TFETs is the possibility of obtaining a 
sub-threshold swing $S$ well below that obtained in conventional
FETs, 
which would in turn allow to strongly reduce the supply voltage for digital logic well below 0.5~V, 
and consequently the power consumption~\cite{Hansch}. 

In this regard, it has been recently observed~\cite{Knoch2} that one-dimensional carbon-based devices 
are more promising than planar silicon devices,
because of the small carrier effective mass, that enhances the $I_{\rm on}$ current, 
and of the small quantum capacitance in the ON state, which guarantees good electrostatic control of 
the channel through the gate voltage, improving $S$.

In this letter, we propose the realization of an ultra-low-power Bilayer Graphene TFET (BG-TFET)
and investigate its performance by means of numerical simulations
based on the self-consistent solution of the 3D Poisson and Schr\"odinger
equations, within the Non-Equilibrium Green's (NEGF), implemented in our
open source code NanoTCAD ViDES~\cite{ViDES}.

We show that a large $I_{\rm on}/I_{\rm off}$ ratio in excess of $10^3$ can be obtained for a voltage 
$V_{\rm DD}$ of only $0.1$~V. 
The low quantum capacitance of bilayer graphene allows us to grab the  advantages provided by one-dimensional 
carbon channels also with a planar --- more easily manufacturable --- structure, and to achieve very low $S$.

\section{Results and Discussion}

The Schr\"odinger equation has been solved within a $p_z$ orbital
basis set in the real space, and details can be found in~\cite{Fiori,IWCE}.

The device structure is depicted in Fig.~\ref{Fig1}a. The considered device is a double-gate BG-TFET, 
driven by two independent gates, biased with $V_{\rm top}$ and $V_{\rm bottom}$, respectively.
 The channel length $L$ is 40 nm, and the
$p^+$ and $n^+$ reservoirs are 40~nm long and doped with molar fraction $f$. The bilayer graphene is embedded
in an SiO$_2$ dielectric, 3~nm-thick, with relative dielectric constant $\varepsilon_r$=3.9.
A gate overlap has also been taken into account, whose effects will be discussed later on.

In Fig.~\ref{Fig1}a a band edge profile of the TFET in the OFF state is also shown. Since the potential in the 
vertical direction falls linearly,
the device is in the OFF state (minimum current in the channel) for 
$V_{\rm bottom}=V_{\rm min} \equiv (V_{\rm top}-V_{\rm bottom})/2+V_{\rm DS}/2$, where
$V_{\rm DS}$ is the drain-to-source voltage.

The conduction and valence bands can be divided
in five sub-regions along the transport direction: the source, the so-called virtual source,
the channel, the virtual drain and the drain. Deep in the source and drain regions 
the conduction and the valence bands coincide because the vertical electric field rapidly goes to zero.
A gap is induced in all the regions surrounded by the top and the bottom gates, where a differential
voltage $V_{\rm diff} \equiv V_{\rm top}-V_{\rm bottom}$ is imposed, i.e. in the virtual source and
drain regions and in the channel.

Six different fluxes of carriers contribute to transport, identified with a letter in Fig.~\ref{Fig1}a: 
tunneling electrons (A and B fluxes),
 tunneling holes (E and D fluxes)
and thermally emitted electrons (C) and holes (F).
In order to obtain a small $I_{\rm off}$,  all mentioned fluxes have to be minimized through proper band engineering. 

One solution could be inducing a large energy gap, which translates in imposing 
a large vertical electric field. This option is however limited by the 
breakdown field of the oxide, and by the gate leakage current.
To this purpose, we have chosen SiO$_2$ as a gate dielectric,
which guarantees an intrinsic breakdown field close to 14 MV/cm~\cite{ianna} (well above the simulated vertical electric fields),
and a large graphene/SiO$_2$ band offset (larger than 3~eV), which ensures very small
gate leakage currents.

The molar fraction of the source and drain leads instead determines $E_1$ and $E_2$, the difference between the mid-gap
potential in the channel ($E_{\rm ref}$) and the mid-gap in the virtual drain and in the drain, respectively (same considerations
obviously follow for the source): the higher $f$, the higher $E_1$ and $E_2$.
Decreasing $E_1$ and $E_2$ can help in reducing E and D (or A and B in the source),
but in turns increases of thermionic emission (C and F).

Let us focus on the performance of the BG-TFET with ultra-low supply voltage  $V_{\rm DD}$=0.1~V.
In Fig.~\ref{Fig1}b we show the transfer characteristics as a function of $V_{\rm bottom}$ of a device with molar 
fraction $f=2.5\times10^{-3}$, for fixed differential gate voltages
$V_{\rm diff}=$6, 6.5, 7 and 8~V. 

As can be noted, differently from what happens in double gate BG-FETs~\cite{Fiori}, 
the BG-TFET can be perfectly switched off, with a steep sub-threshold behavior (always smaller than 20~mV/dec)
and a large $I_{\rm on}/I_{\rm off}$ ratio also for a very low $V_{\rm DD}$ (e.g. $I_{\rm on}/I_{\rm off}$=4888 
for $V_{\rm diff}$=8~V).

Due to the large vertical electric field, particular attention has also to be posed on the gate leakage current. 
We can estimate its value using a well-tested analytical model for Si-SiO$_2$ gate stacks based on WKB approximation of 
the triangular barrier~\cite{Depas}. We find that the gate current is of the order of $10^{-6}$ $\mu$A/$\mu$m
for $V_{\rm diff}=7$~V, and $10^{-4}$ $\mu$A/$\mu$m for $V_{\rm diff}=8$~V, i.e. negligible with respect to the smallest 
drain currents considered in our simulations.
%


In Fig.~\ref{Fig2}a we show the transfer characteristics as a function of $V_{\rm bottom}$, for $V_{\rm diff}$=7~V, and
for different dopant molar fractions $f$ in the drain and source leads. As $f$ 
is increased, the $I_{\rm on}/I_{\rm off}$ ratio
degrades, since $I_{\rm off}$ increases, while $I_{\rm on}$ remains almost constant.

In Fig.~\ref{Fig2}b, we show the transfer characteristics as a function of gate overlap. As can be
seen, the smaller the overlap, the higher the $I_{\rm off}$ current,
since it is the energy gap of the virtual reservoirs that suppresses tunneling from contacts to the channel.

In order to investigate the performance of single-gate devices, we have also performed
simulations of a device with asymmetric gate oxide thicknesses (Fig.~\ref{Fig2}c): top oxide is 3~nm, and the bottom oxide is 9~nm. 
In this case the bottom gate voltage is kept at a constant bias of $-12$~V, whereas
the top gate is used as the control gate.

As can be noted, an $I_{\rm on}/I_{\rm off}$ ratio 
of few hundreds can be still obtained. The sub-threshold swing (and then the $I_{\rm on}/I_{\rm off}$ ratio)
can be improved with a thicker bottom oxide of the order of 100 nm, that would reduce the capacitive coupling 
between channel and bottom gate. We however
limited our analysis to 9~nm, because of numerical convergence problems encountered when dealing with larger structures.

As noted in Ref.~\cite{Knoch2}, one-dimensional channels
are typically more appealing for tunneling device applications than planar two-dimensional channels,
since their quantum capacitance $C_Q$ is smaller than the electrostatic capacitance
$C_{\rm ox}$ even in the ON state. Such condition guarantees a good gate control over the potential barrier 
 and larger $I_{\rm on}/I_{\rm off}$. 

In Fig.~\ref{Fig3}a, we show $\phi_C$, the electrostatic potential in the middle of the channel,
 as a function of the bottom gate voltage, for $f=2.5\times10^{-3}$ and
$V_{\rm diff}=7$~V. On the same plot, we also show the line $\phi_C=V_{\rm bottom}-V_{\rm min}+V_{\rm DS}/2$, corresponding to ideal control of the channel potential from the gate voltage.
As can be seen, also for larger gate voltages, $\phi_C$ reasonably follows $V_{\rm bottom}$, and no channel potential saturation is observed. 
As indeed shown in Fig.~\ref{Fig3}b, where the ratio $C_{\rm Q}/C_{\rm ox}$ 
is depicted, for small voltages the device is working in the Quantum Capacitance Limit (i.e. $C_{\rm Q} \ll C_{\rm ox}$), whereas for 
larger values $C_Q$ is comparable with $C_{\rm ox}$. However, it never reaches the condition
$C_{\rm Q} \gg C_{\rm ox}$ as in silicon TFETs. As a consequence, despite its bidimensionality, bilayer 
graphene is appealing for tunneling devices applications.

\section{Conclusion}

In conclusions, we have performed a numerical analysis of a bilayer 
graphene tunnel field effect transistors, based on the self-consistent solution of the 3D Poisson/Schr\"odinger
equations, within the NEGF formalism.
Large $I_{\rm on}/I_{\rm off}$ ratios can be obtained even for an ultra-low supply voltage of only 100~mV, 
thanks to an extremely steep sub-threshold slope.
The low quantum capacitance of bilayer graphene allows the BG-TFET to have most of the 
advantages of one-dimensional TFETs, but none of the disadvantages in terms of structure patterning and 
lithography. Indeed, since present-day lithography is adequate for the fabrication of BG-TFETs, and the 
single-gate driving option with constant bottom gate bias is suitable for planar integration, 
we consider the device very promising for fabrication experiments and circuit integration.

\section*{Acknowledgment}

The work was supported in part by the EC Seventh Framework
Program under project GRAND (Contract 215752), by the Network of Excellence
NANOSIL (Contract 216171), and by the European Science Foundation
EUROCORES Program Fundamentals of Nanoelectronics, through funds from
CNR and the EC Sixth Framework Program, under project DEWINT (ContractERAS-CT-2003-980409).

\newpage

\bibliography{fiori}

\newpage


\begin{figure}
\caption{a) Sketch of the double-gate bilayer graphene TFET: the channel length is 40~nm, and n$^+$ and p$^+$ 
reservoirs are 40~nm long with molar fraction $f$. The device is embedded in 3~nm-thick SiO$_2$ dielectric. $V_{\rm top}$ and $V_{\rm bottom}$ 
are the voltages applied the top and bottom gate, respectively. Gate overlap has been also considered. Below, 
band edge profile of the device in the OFF state; b) Transfer characteristics of the double-gate BG-TFET
for different $V_{\rm diff}$. $f$ is equal to $2.5\times10^{-3}$ and $V_{\rm DS}=0.1$~V.}
\label{Fig1}	
\caption{a) Transfer characteristics of TBG-FETs for $V_{\rm diff}$=7~V, $V_{\rm DS}=0.1$~V, 
and different molar fractions $f$. $V_{\rm min}$=-3.55~V;
 b) Transfer characteristics of BG-TFET for $V_{\rm diff}$=7~V, $V_{\rm DS}=0.1$~V, and different gate overlap. 
$V_{\rm min}$=-3.55~V.
c) Transfer characteristic for a device with asymmetric oxide thicknesses: the top oxide thickness 
is 3~nm, and the bottom oxide thickness is 9~nm. $V_{\rm bottom}$ is fixed to -12~V and $V_{\rm DS}=0.1$~V.}
\label{Fig2}	
\caption{a) electrostatic potential in the middle of the channel $\phi_C$ as a function of the bottom
gate voltage (solid line). The dashed line refers the case in which the gate control is ideal, i.e.
$\phi_c=(V_{\rm bottom}-V_{\rm min})+V_{\rm DS}/2$. In the inset the equivalent capacitance circuit 
of the device is shown;
b) ratio of the quantum capacitance $C_Q$ and the oxide capacitance $C_{\rm ox}$.}
\label{Fig3}
\end{figure}


\begin{figure}
\begin{center}
\vspace{3cm}
\epsfig{file=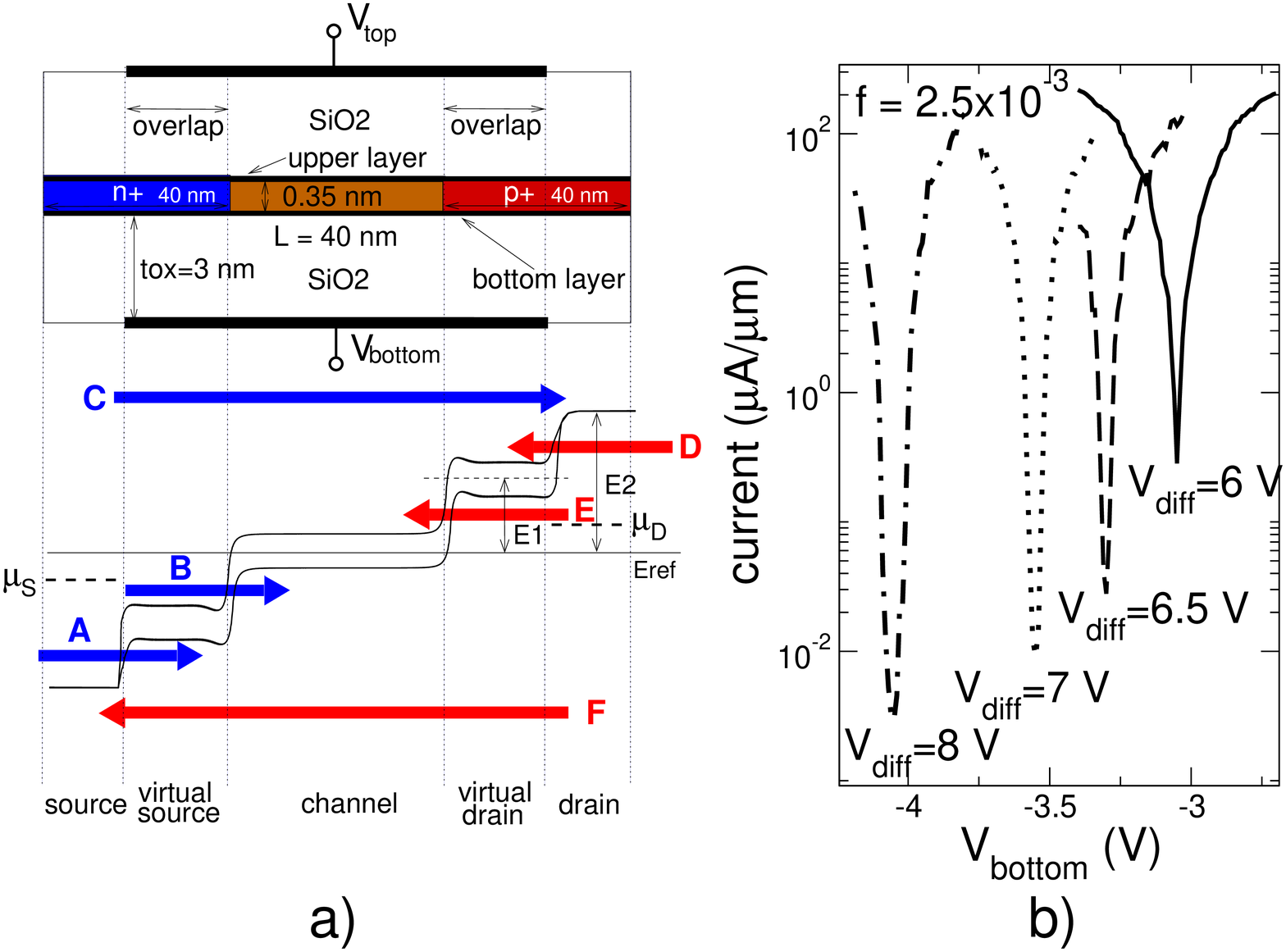,width=17cm} 
\huge

FIG. 1\\
\normalsize	
\end{center}
\end{figure}

\begin{figure}
\begin{center}
\epsfig{file=./Fig2.eps,width=17cm} 
\huge
\vspace{1cm}\\
FIG. 2\\
\normalsize	
\end{center}
\end{figure}

\begin{figure}
\begin{center}
\epsfig{file=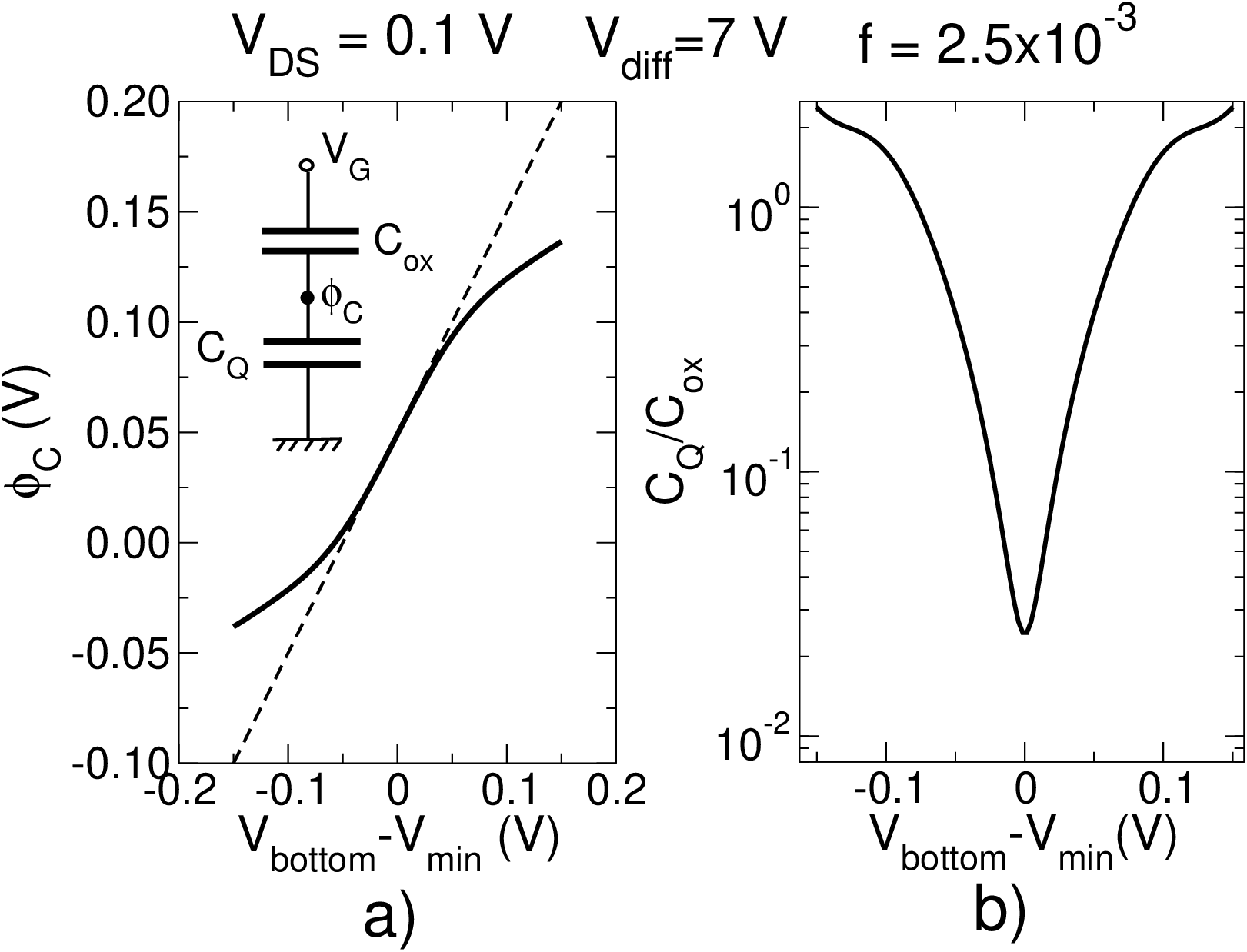,width=17cm} 
\huge
\vspace{1cm}\\
FIG. 3\\
\normalsize	
\end{center}
\end{figure}

\end{document}